\documentclass[varenna]{cimento}
\usepackage{graphicx}  
\newcommand{\gapprox}{\stackrel{>}{\sim}}
\newcommand{\lapprox}{\stackrel{<}{\sim}}
\title{\bf Exact treatment of trapped imbalanced fermions in the BEC limit}
\author{P. Pieri and G.C. Strinati}
\institute{Dipartimento di Fisica,  Universit\`{a} di Camerino, 
I-62032 Camerino, Italy}

\begin{document}

\maketitle

\section{Introduction}
In the present contribution we shall analyze the effects of imbalancing the populations of two-component 
trapped fermions in the BEC (strong-coupling) limit of the attractive interaction between fermions 
of different components.
In particular, we shall derive a set of coupled equations which describe composite bosons and excess 
fermions in this limit, starting from the gap equation with two different fermionic chemical potentials.
Care will be used to include in these equations the processes leading to the correct dimer-dimer 
and dimer-fermion scattering lengths, which require us to consider beyond-mean-field effects.
Numerical results will be presented for the density profiles of composite bosons and excess fermions, 
which are relevant to the recent experiments with trapped Fermi atoms.
Results for the formation of vortex patterns in the presence of density imbalance will also be presented.

The interest in imbalanced populations of fermions with different spins originated long time ago from
condensed-matter physics, where a magnetic field can alter the populations of spin $\uparrow$ and 
$\downarrow$ electrons \cite{Clogston-1962,Sarma-1963}.

Quite generally, density imbalance $\delta n = n_{\uparrow} - n_{\downarrow}$ between ``spin up'' and
``spin down'' fermions introduces \emph{a new degree of freedom} in the system, and may thus possibly 
lead to the removal degeneracies and the occurrence of novel phases \cite{FF-1964,LO-1965}. 

In this context, trapped cold Fermi atoms offer a unique possibility for observing the consequences 
of density imbalance.
These systems has lately be object of mounting interest, both experimentally and theoretically, as
they allow one to explore the BCS-BEC crossover by controlling the interaction between fermions 
of different components via the use of Fano-Feshbach resonances.
Recently, two experimental studies with imbalanced populations have raised novel interest in these 
systems \cite{Ketterle-2006,Hulet-2006}.
Density profiles of the two fermionic species as well as vortices have been detected.
A quantum phase transition to the normal state on the BCS side of the crossover 
as well as phase separation in the crossover region have been identified.

From a theoretical point of view, the BCS-BEC crossover gets modified as the many-body problem becomes
richer in the presence of density imbalance.
The interest in this problem has then involved not only cold-atom and condensed-matter physics, but 
also nuclear and subnuclear physics.
On the BEC side of the crossover, it further gives one the opportunity of embedding into the diagrammatic
structure the processes \cite{KC-2005,BK-1998} leading to the correct values of the scattering lengths 
for two composite bosons ($a_{{\rm B}}$) and for a composite boson with an excess fermion ($a_{{\rm BF}}$).

The effects of density imbalance on fermionic superfluids were originally studied in the weak-coupling (BCS) 
limit of the crossover both for the homogeneous \cite{BCS-d-i-homogeneous} and 
trapped case \cite{machida}.
Only recently these calculations have been extended to cover the BCS-BEC crossover 
\cite{BCS-BEC-d-i-homogeneous}, and to consider the effects of the trap 
\cite{PS-2006,theoretical-density-imbalance-trap,Duan-Yip-2006} which are essential to account 
for the experimental results with density imbalance.
We thus begin by giving in the next Section a brief account of the mean-field approach for the homogenous
case when different spin populations are considered, with emphasis to the strong-coupling (BEC) limit to 
which we shall eventually restrict in the later Sections when considering the trapped case.

\vspace{0.5cm}  
\section{Mean-field treatment for the homogeneous case}

The microscopic BCS theory of superconductivity is most conveniently formulated in terms of fermionic
single-particle Green's functions \cite{FW}.
Due to the presence of spontaneous broken symmetry, anomalous averages need be considered together with
normal averages, leading to a $2 \times 2$ matrix for the Green's functions.

This formulation can be readily extended to include different populations for the two fermionic species 
(labeled by spin-$\uparrow$ and spin-$\downarrow$) which mutually interact via a contact potential.
This is done by considering two different chemical potentials $\mu_{\uparrow}$ and $\mu_{\downarrow}$ 
for the two species, so that the \emph{equation of motion} for the fermionic single-particle Green's 
functions $G_{ij}(\mathbf{k},\omega_{s})$ within mean field reads:
\begin{equation}
\left[ \begin{array}{cc}
i \omega_{s} - \frac{\mathbf{k}^2}{2 m} + \mu_{\uparrow}   &  - \Delta  \\
- \Delta^*   &  i \omega_{s} + \frac{\mathbf{k}^2}{2 m} - \mu_{\downarrow} \end{array} \right] 
\left[ \begin{array}{cc}
G_{11}  &  G_{12} \\
G_{21}  &  G_{22}\end{array} \right] \, = \,  
\left[ \begin{array}{cc}
1  &  0 \\
0  &  1 \end{array} \right]  \,\, .                           \label{Greeen-function-mean-field}                                                           
\end{equation}
Here, $\omega_{s} = (2s+1)\pi/(k_{{\rm B}} T)$ ($s$ integer) is a fermionic Matsubara frequency at
temperature $T$, $\mathbf{k}$ a wave vector, $m$ the fermion mass, and $\Delta$ the gap function.
The novelty introduced in eq.~(\ref{Greeen-function-mean-field}) by population imbalance 
is the presence of two different chemical potentials in the diagonal
matrix elements on its left-hand side.
This seemingly minor difference will, however, yield significant different results 
upon inverting eq.~(\ref{Greeen-function-mean-field}) in favor of $G_{ij}$.

From this inversion one obtains, in particular: 
\begin{eqnarray}
n & = & n_{\uparrow} \, + n_{\downarrow} \, =  \sum_{k} \, 
\left[ e^{i\omega_{s} \eta} G_{11}(k) \, - \, e^{-i\omega_{s} \eta} G_{22}(k) \right] \nonumber \\ 
& = & \int \! \, \frac{d\mathbf{k}}{(2 \pi)^{3}} 
\left\{ 1 - \frac{\xi(\mathbf{k})}{E(\mathbf{k})} \left[ 1 - f(E_{+}(\mathbf{k})) 
- f(E_{-}(\mathbf{k})) \right] \right\}                         \label{total-density-homogeneous}
\end{eqnarray}
for the total density (where $\eta$ is a positive infinitesimal),
\begin{eqnarray}
\delta n & = & n_{\uparrow} \, - n_{\downarrow} \, =  \sum_{k} \, 
\left[ e^{i\omega_{s} \eta} G_{11}(k) \, + \, e^{-i\omega_{s} \eta} G_{22}(k) \right] \nonumber \\ 
& = & \int \! \, \frac{d\mathbf{k}}{(2 \pi)^{3}} 
\left\{ f(E_{+}(\mathbf{k})) - f(E_{-}(\mathbf{k})) \right\}  \label{difference-density-homogeneous}
\end{eqnarray}
for the density difference, and
\begin{equation}
- \frac{m}{4 \pi a_{{\rm F}}} \, = \, \int \! \, \frac{d\mathbf{k}}{(2 \pi)^{3}}
\left\{ \frac{ 1 - f(E_{+}(\mathbf{k})) - f(E_{-}(\mathbf{k}))}{2 E(\mathbf{k})} \, -
\, \frac{m}{\mathbf{k}^{2}}  \right\}                                   \label{regularized-gap-equation}
\end{equation}
for the gap equation where the strength of the contact fermionic attraction has been replaced by the
fermionic scattering length $a_{{\rm F}}$ by suitable regularization \cite{PS-2000}.
In these expressions, $f(E) = ( e^{\beta E} + 1 )^{-1}$ is the Fermi distribution function with 
$\beta=1/(k_{{\rm B}} T)$ and we have introduced the notation:
\begin{equation}
E_{\pm}(\mathbf{k}) \, = E(\mathbf{k}) \, \pm \delta \xi \,\,\,\,\,\,\,\,\,  , \,\,\,\,\,\,\,\,\,
E(\mathbf{k}) \, = \, \sqrt{ \xi(\mathbf{k})^{2} \, + \, |\Delta|^{2} }      \label{notation-I}
\end{equation}
and
\begin{equation}
\xi(\mathbf{k}) \, = \, \frac{\mathbf{k}^2}{2 m} \, - \, \frac{\mu_{\uparrow} + \mu_{\downarrow}}{2} 
\,\,\,\,\,\,\,\,\,  , \,\,\,\,\,\,\,\,\, \delta \xi \, = \, \frac{\mu_{\downarrow} - \mu_{\uparrow}}{2}
                                                                              \label{notation-II}
\end{equation}
[only the case of equal fermion masses will be considered throughout].
Note from eq.~(\ref{difference-density-homogeneous}) that, in the low temperature limit, the only way
to sustain a nonvanishing value of $\delta n$ is to have either $E_{+}(\mathbf{k})$ or 
$E_{-}(\mathbf{k})$ negative.
For definiteness, we shall assume $n_{\uparrow} \ge n_{\downarrow}$.

It is known from the theory of the BCS-BEC crossover that the dimensionless parameter $(k_{{\rm F}}\, a_{{\rm F}})^{-1}$
controls the evolution 
from the weak-coupling BCS regime (where $a_{{\rm F}} < 0$ and $(k_{{\rm F}}\, a_{{\rm F}})^{-1} \lapprox -1$) 
to the strong-coupling BEC regime (where $a_{{\rm F}} > 0$ and $(k_{{\rm F}}\, a_{{\rm F}})^{-1} \gapprox +1$), 
with the ``crossover'' region  being limited in practice to the interval 
$-1 \lapprox (k_{{\rm F}}\, a_{{\rm F}})^{-1} \lapprox +1$.
%
%
%
Here, $k_{{\rm F}}$ is the Fermi wave vector related in the standard way to the total density.

In the following, we shall mostly be interested in the strong-coupling (BEC) regime at low temperature.
In this limit, one expects the presence of density imbalance to produce a density $n_{\downarrow}$ of
composite bosons formed by pairing a fermion of spin ${\uparrow}$ with a fermion of spin $\downarrow$,
plus a density $\delta n = n_{\uparrow} - n_{\downarrow}$ of excess fermions of spin $\uparrow$.
Correspondingly, one finds from eqs.~(\ref{total-density-homogeneous})-(\ref{difference-density-homogeneous}): 
\begin{equation}
\mu_{\downarrow} \, = \, - \, \epsilon_{0} \, - \, \mu_{\uparrow} \, + \, \mu_{{\rm B}}         \label{mu-down}
\end{equation}
where $\epsilon_{0} \, = \, \frac{1}{m \, a_{{\rm F}}^{2}}$ is the binding energy of the associated two-body
problem and
\begin{equation}
\mu_{\uparrow} \, \cong \, \epsilon_{{\rm F}}(\delta n) \, + \, 
\left( \frac{2 \pi a_{{\rm BF}}}{m_{{\rm BF}}} \right) \, n_{0}                                       \label{mu-up}
\end{equation}

\begin{equation}
\mu_{{\rm B}} \, \cong \, \left( \frac{4 \pi a_{{\rm B}}}{m_{{\rm B}}} \right) \, n_{0}
\, + \, \left( \frac{2 \pi a_{{\rm BF}}}{m_{{\rm BF}}} \right) \, \delta n   \,\, .                   \label{mu-B}
\end{equation}
In these expressions, $\epsilon_{{\rm F}}(\delta n)$ is the Fermi energy corresponding to $\delta n$,
$\mu_{{\rm B}}$ plays the role of the chemical potential of the composite bosons, and $n_{0}$  is the
condensate density which equals $(n - \delta n)/2$ under the present circumstances. 
In addition, $m_{{\rm B}} = 2 m$ is the mass of a composite boson, $m_{{\rm BF}} = \frac{2}{3} m$ is the reduced mass 
of the two-body system made up of a composite boson and an excess fermion, while $a_{{\rm B}} = 2 a_{{\rm F}}$ is the 
scattering length for the scattering of two composite bosons and $a_{{\rm BF}} = (8/3) a_{{\rm F}}$ 
for the scattering of a composite boson and an excess fermion.
These values for $a_{{\rm B}}$ and $a_{{\rm BF}}$ are specific to the mean-field treatment and correspond to the
Born approximation for the scattering processes.

After these preliminary considerations of general relevance in the presence of imbalanced fermion populations, 
we pass now to consider the trapped case which is specifically relevant to the experiments with cold atoms.

\section{Mean-field treatment for the trapped case}

At a formal level the mean-field treatment for the inhomogeneous case proceeds along similar lines as for 
the homogenous case discussed in the previous Section, the difference being that the \emph{equation of 
motion} for the fermionic single-particle Green's functions $G_{ij}(\mathbf{r},\mathbf{r}';\omega_{s})$ 
is now given by:
\begin{equation}
\left[ \begin{array}{cc}
i \omega_{s} - \mathcal{H}_{\uparrow}(\mathbf{r})  &  - \Delta(\mathbf{r})  \\
- \Delta(\mathbf{r})^*   &  i \omega_{s} + \mathcal{H}_{\downarrow}(\mathbf{r}) \end{array} \right] 
\left[ \begin{array}{cc}
G_{11}  &  G_{12} \\
G_{21}  &  G_{22}\end{array} \right] \, = \,  
\delta(\mathbf{r} - \mathbf{r}') 
\left[ \begin{array}{cc}
1  &  0 \\
0  &  1 \end{array} \right]  \,\, .                         \label{Greeen-function-mean-field-trapped}                                                         
\end{equation}
Here, $\mathcal{H}_{\sigma}(\mathbf{r}) = - \frac{\nabla^{2}}{2 m} + V(\mathbf{r}) - \mu_{\sigma}$ is the 
single-particle Hamiltonian in the presence of the trapping potential $V(\mathbf{r})$.
Equation (\ref{Greeen-function-mean-field-trapped}) is the Green's function version of the 
Bogoliubov-de Gennes equations \cite{BdG}, which are often used to describe inhomogeneous superconductors.

As in the previous Section, we are mostly interested in the strong-coupling BEC regime at low temperature.
In the case of equal fermion populations (whereby $\mu_{\uparrow} = \mu_{\downarrow}$), it has been shown
\cite{PS-2003} that the Gross-Pitaevskii equation for composite bosons can be derived in this limit from 
the Bogoliubov-de Gennes equations (\ref{Greeen-function-mean-field-trapped}).
This mapping enables one to exploit the results obtained directly from the more manageable bosonic 
Gross-Pitaevskii equation and, when needed, to use them as benchmarks for the fermionic calculation 
in the limit.
This analysis has recently been extended to the imbalanced case when $\mu_{\uparrow} \ne \mu_{\downarrow}$.
In this case, in the place of the Gross-Pitaevskii equation one ends up with two coupled equations that 
describe the simultaneous presence of composite bosons and excess fermions \cite{PS-2006}.
As the analysis proceeds along similar lines in both (balanced and imbalanced) cases, we indicate it here
schematically for convenience.

One starts by rewriting the Bogoliubov-de Gennes equations (\ref{Greeen-function-mean-field-trapped}) 
in the integral form:
\begin{equation}
\hat{\mathcal{G}}(\mathbf{r},\mathbf{r}';\omega_{s}) =  \hat{\mathcal{G}_{0}}(\mathbf{r},\mathbf{r}';\omega_{s}) 
+  \int \! d\mathbf{r}'' \, \hat{\mathcal{G}_{0}}(\mathbf{r},\mathbf{r}'';\omega_{s}) \, \hat{B}(\mathbf{r}'')
\, \hat{\mathcal{G}}(\mathbf{r}'',\mathbf{r}';\omega_{s})                                           \label{B-dG-integral-form}                       
\end{equation}
where
\begin{equation}
\hat{\mathcal{G}_{0}}(\mathbf{r},\mathbf{r}';\omega_{s}) \, = \, 
\left[ \begin{array}{cc}
\tilde{\mathcal{G}_{0}}(\mathbf{r},\mathbf{r}';\omega_{s}|\mu_{\uparrow})   &   0  \\
0   &  - \, \tilde{\mathcal{G}_{0}}(\mathbf{r}',\mathbf{r};-\omega_{s}|\mu_{\downarrow})   \end{array} 
\right]                                                          \label{Greeen-function-mean-non-interacting}                                        
\end{equation}
is the matrix of the noninteracting Green's functions which satisfy the equation
\begin{equation}
\left[ i \omega_{s} \, - \, \mathcal{H}_{\sigma}(\mathbf{r}) \right] \,
\tilde{\mathcal{G}_{0}}(\mathbf{r},\mathbf{r}';\omega_{s}|\mu_{\sigma}) \, = \, \delta(\mathbf{r} - \mathbf{r}'),
\end{equation}
being subject to the same trapping potential $V(\mathbf{r})$ entering the single-particle Hamiltonian
$\mathcal{H}_{\sigma}(\mathbf{r})$.
In addition, the matrix
\begin{equation}
\hat{B}(\mathbf{r}) \, = \, 
\left[ \begin{array}{cc}
           0           &   \Delta(\mathbf{r})  \\
\Delta^{*}(\mathbf{r}) &            0              \end{array}  \right]           \label{B-matrix}                                                                 
\end{equation}
contains the effects of the interaction via the gap function.

At this point, expansion of $\mathcal{G}_{11}(\mathbf{r},\mathbf{r}';\omega_{s})$ and 
$\mathcal{G}_{22}(\mathbf{r},\mathbf{r}';\omega_{s})$ up to order $\Delta^{2}$ yields the expressions for the
local densities $n_{\uparrow}(\mathbf{r})$ and $n_{\downarrow}(\mathbf{r})$, in the order, while expansion of
$\mathcal{G}_{12}(\mathbf{r},\mathbf{r}';\omega_{s})$ up to order $\Delta^{3}$ yields an equation for the
local gap $\Delta(\mathbf{r})$.
The gap equation can be cast in the form \cite{PS-2006}:
\begin{equation}
- \frac{\nabla^{2}}{2 m_{{\rm B}}} \Phi(\mathbf{r}) 
+  \left[ 2 V(\mathbf{r}) + \frac{3 \pi a_{{\rm BF}}}{m} \, \delta n(\mathbf{r}) \right] \Phi(\mathbf{r})              
+  \frac{4 \, \pi \, a_{{\rm B}}}{m_{{\rm B}}} \, |\Phi(\mathbf{r})|^{2} \, \Phi(\mathbf{r}) = \mu_{{\rm B}} \, \Phi(\mathbf{r})     \label{local-gap-equation}   
\end{equation}
where again $m_{{\rm B}} = 2 m$, $a_{{\rm B}} = 2 a_{{\rm F}}$, $a_{{\rm BF}} = \frac{8}{3} a_{{\rm F}}$,
$\mu_{{\rm B}} = \mu_{\uparrow} + \mu_{\downarrow} + \epsilon_{0}$, while
$\Phi(\mathbf{r}) = \sqrt{\frac{m^{2} a_{{\rm F}}}{8 \pi}} \, \Delta(\mathbf{r})$ 
plays the role of the bosonic \emph{condensate wave function}.

By a similar token, the density of excess fermions reads:
\begin{equation}
\delta n(\mathbf{r}) =  n_{\uparrow}(\mathbf{r}) \, - \, n_{\downarrow}(\mathbf{r})                                
\cong \int \! \, \frac{d\mathbf{k}}{(2 \pi)^{3}} \, 
f \left(\frac{\mathbf{k}^{2}}{2m}+V(\mathbf{r})+\frac{3\pi a_{{\rm BF}}}{m} |\Phi(\mathbf{r})|^{2}-\mu_{\uparrow}\right) \,\, . 
                                                                                             \label{density-excess-fermions}
\end{equation}
Note that the two coupled equations (\ref{local-gap-equation}) and (\ref{density-excess-fermions}) embody the mutual 
effects of the bosonic distribution $|\Phi(\mathbf{r})|^{2}$ and the fermionic distribution $\delta n(\mathbf{r})$. 
Finally, the expression for the density is:  
\begin{equation}
n(\mathbf{r}) \, = \, n_{\uparrow}(\mathbf{r}) \, + \, n_{\downarrow}(\mathbf{r}) \, = \, 
\delta n(\mathbf{r}) + 2 \, |\Phi(\mathbf{r})|^{2}  \label{total-density}                               
\end{equation}
which completes the set of three coupled equations for $\Phi(\mathbf{r})$, $\delta n(\mathbf{r})$, and $n(\mathbf{r})$. 

The remaining problem is that the scattering between composite bosons (as embodied by $a_{{\rm B}}$) and between a composite 
boson and an excess fermion (as embodied by $a_{{\rm BF}}$) has been treated so far at the lowest order within the Born 
approximation, corresponding to the values $a_{{\rm B}} = 2 a_{{\rm F}}$ and $a_{{\rm BF}} = \frac{8}{3} a_{{\rm F}}$ obtained by our
derivation.
This points to the need of going beyond the mean-field treatment to include the full set of scattering processes 
for $a_{{\rm B}}$ and $a_{{\rm BF}}$.
In diagrammatic language, this implies identifying additional fermionic diagrams containing diagrammatic sequences 
for $a_{{\rm B}}$ and $a_{{\rm BF}}$ as sub-units.
\section{Exact equations in the dilute case}

The validity of the equations obtained in the previous Section can be extended by improving on the values of the 
scattering lengths $a_{{\rm B}}$ and $a_{{\rm BF}}$.
From the exact solutions (in real-space representation) of the three- \cite{STM-1957} and four-body \cite{PSS-2004} 
problems it is known that the correct values are $a_{{\rm BF}} = 1.18 a_{{\rm F}}$ and $a_{{\rm B}}= 0.6 a_{{\rm F}}$, 
in the order.
These values have also been determined \emph{diagrammatically} (in wave-vector representation) in the limit of 
vanishing density, for $a_{{\rm BF}}$ in ref.~\cite{BK-1998} and for $a_{{\rm B}}$ in ref.~\cite{KC-2005}.
To improve on the derivation of the equations obtained in the previous Section, one has thus to embed the 
diagrammatic sub-units, which identify $a_{{\rm B}}$ and $a_{{\rm BF}}$ in the limit of vanishing density, 
into the many-body structure at finite density.
This can be achieved as follows.

\begin{figure}
\begin{center}
\includegraphics*[width=7.5cm]{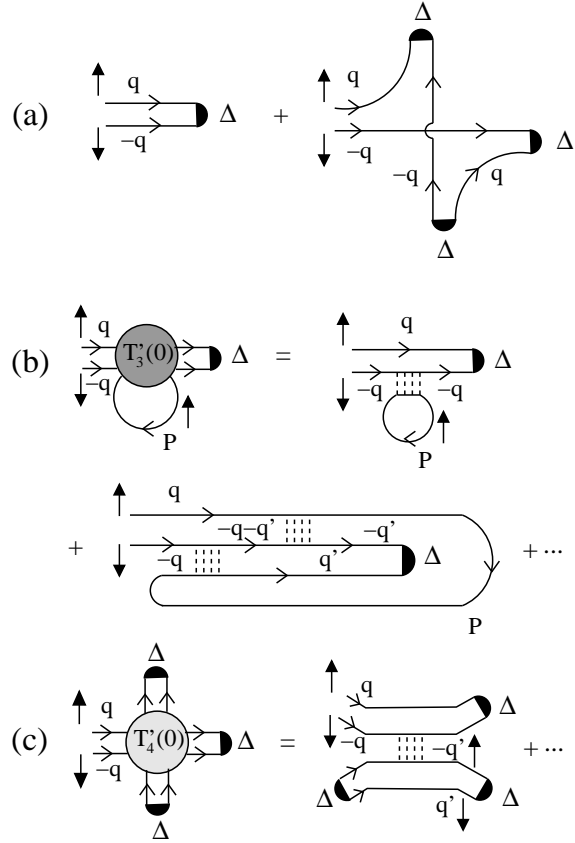}
\caption{Tadpole diagrams for composite bosons describing the gap equation in the BEC limit.
                   (a) Mean-field contributions; 
                   Contributions beyond the Born approximation which include the scattering processes yielding
                   the correct values of the (b) dimer-fermion and (c) dimer-dimer scattering lengths.
                   Full lines represent the fermion propagators of given spin and broken lines represent the 
                   bare fermionic interaction. 
                   [Reproduced from ref.~\cite{PS-2006}.]} 
\end{center}
\end{figure} 

As a natural extension of what is done for determining the condensate density of point-like bosons \cite{Popov},
in the BEC limit of interest here the gap equation can be interpreted as the condition of vanishing ``tadpole'' 
insertions for composite bosons.
At the mean-field level considered in the previous Section, the diagrams representing this condition are depicted 
in fig.~1(a), with the understanding that a composite-boson propagator with zero four-momentum can be inserted 
from the left while the gap $\Delta$ corresponds to a condensate line.
These diagrams account for the dimer-fermion and dimer-dimer scattering processes within the Born approximation.
The additional tadpole diagrams, which yield the correct values of $a_{{\rm BF}}$ and $a_{{\rm B}}$, are depicted 
in figs.~1(b) and 1(c), in the order.
Here, the dimer-fermion and dimer-dimer scattering processes exclude the Born-approximation contributions already 
included at the mean-field level in fig.~1(a). 

With this procedure, one ends up formally with the same coupled equations (\ref{local-gap-equation})-(\ref{total-density})
which, however, now contain the correct values of $a_{{\rm B}}$ and $a_{{\rm BF}}$.
In this context, a comment is worth on how to factor out from the diagrams of fig.~1(b) the product $a_{{\rm BF}}$ 
times $\delta n$ that enters eq.~(\ref{local-gap-equation}).
The point is that the integration over the wave vector $\mathbf{P}$ is bounded within the Fermi sphere of radius 
$\sqrt{2 m \mu_{\uparrow}}$, while the remaining integrations over $\mathbf{q}$, $\mathbf{q'}$, $\cdots$, extend 
outside this Fermi sphere.
One may, accordingly, neglect the $P$-dependence everywhere in the diagrams of fig.~1(b) \emph{except\/} in the
fermion propagator labeled by $P$ and corresponding to spin-$\uparrow$ fermions.
The density of excess fermions results in this way from the $P$-integration, with the remaining parts
of the diagrams yielding the exact dimer-fermion scattering matrix $T^{'}_{3}(0)$.
This excludes, by definition, the Born contribution resulting from mean field.
A similar analysis can be carried out for the density equation (\ref{density-excess-fermions}), where only the 
value of the dimer-fermion scattering length $a_{{\rm BF}}$ requires corrections beyond mean field \cite{PS-2006}.

\section{Numerical results and comparison with experiments}

We pass now to report on the numerical solution of the coupled equations (\ref{local-gap-equation})-(\ref{total-density})
with the exact values of $a_{{\rm B}}$ and $a_{{\rm BF}}$.
By their very derivation, these equations are expected to be valid in the strong-coupling (BEC) regime where $a_{{\rm F}} > 0$,
which however extends as down as to $(k_{{\rm F}}\, a_{{\rm F}})^{-1} \approx +1$ for all practical purposes.

We determine the density profiles $\delta n(\mathbf{r})$ and $n_{0}(\mathbf{r}) = |\Phi(\mathbf{r})|^{2}$ in the Thomas-Fermi 
(LDA) approximation (which corresponds to neglecting the kinetic energy term in eq.~(\ref{local-gap-equation})) for a spherical 
trap, as functions of the \emph{asymmetry parameter} 
$\alpha = (N_{\uparrow} \, - \, N_{\downarrow})/(N_{\uparrow} \, + \, N_{\downarrow})$
(where $0 \le \alpha \le 1$).
Numerical results \cite{PS-2006} are shown in:

\begin{figure}
\begin{center}
\includegraphics*[width=6.5cm]{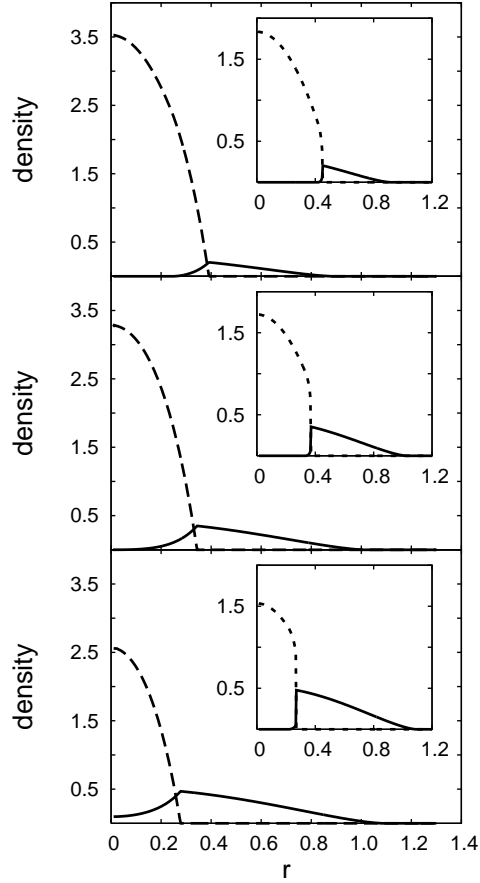}
\caption{Density profiles $\delta n(r)$ (full lines) and $n_{0}(r)$ (broken lines) vs.~$r$
                   for $(k_{{\rm F}} a_{ F})^{-1}=3$ and $\alpha = 0.2$ (upper panel), $\alpha = 0.5$ (middle panel), 
                   $\alpha = 0.8$ (lower panel).
                   The insets show the results when $(k_{{\rm F}} a_{{\rm F}})^{-1}=1$ for the same values of $\alpha$.
                   Here, $r$ is in units of $R_{{\rm TF}}$ and densities are in units of $(N_{\uparrow} + N_{\downarrow}) / R_{{\rm TF}}^{3}$.
                   The Thomas-Fermi radius $R_{{\rm TF}}$ and the Fermi wave vector $k_{{\rm F}}$ are defined for equal populations.
                   [Reproduced from ref.~\cite{PS-2006}.]}
\end{center}
\end{figure} 

(i) fig.~2 for the density profiles $\delta n(r)$ and $n_{0}(r)$ vs. the distance $r = |\mathbf{r}|$ from 
    the center of the trap, for three characteristic values of $\alpha$ and for the coupling $(k_{{\rm F}} a_{{\rm F}})^{-1}=3$ on 
    the BEC side of the crossover.
    The insets show the results for the smaller coupling $(k_{{\rm F}} a_{{\rm F}})^{-1}=1$.
    A spatial separation results between the condensed composite bosons and the excess fermions, which appears sharper 
    for the smaller coupling, corresponding to enhanced effects of the dimer-fermion repulsion.
    Upon approaching  unitarity [$(k_{{\rm F}} a_{{\rm F}})^{-1} = 0$], there thus appears a tendency toward \emph{phase separation}
    with a superconducting core of fully paired fermions surrounded by a cloud of excess fermions.
    For the couplings here considered, this phase separation occurs for all values of $\alpha$.
    Note that, for each value of $\alpha$, the maximum of $\delta n(r)$ occurs where $n_{0}(r)$ vanishes, and that
    there occurs a progressive size shrinking of $n_{0}(r)$ for increasing $\alpha$, with a simultaneous penetration 
    of $\delta n(r)$ toward the center of the trap.
\begin{figure}
\begin{center}
\includegraphics*[width=6.5cm]{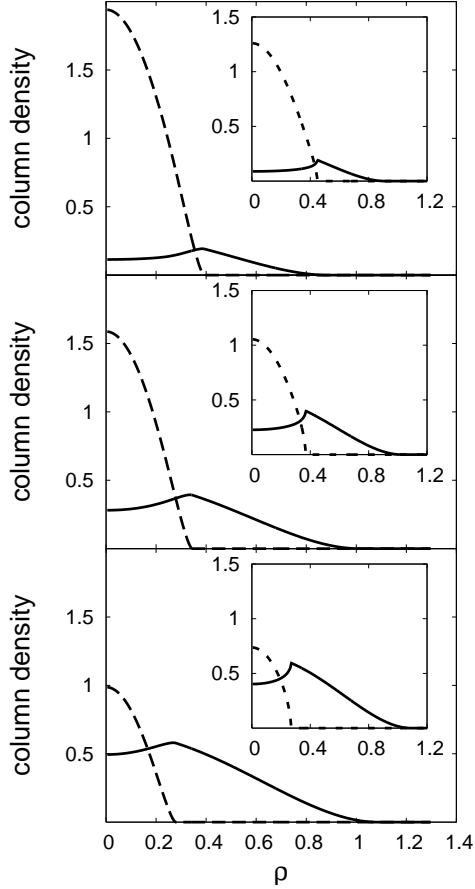}
\caption{Column density profiles $\delta n(\rho)$ (full lines) and $n_{0}(\rho)$ (broken lines) vs.~$\rho$
                   for $(k_{{\rm F}} a_{{\rm F}})^{-1}=3$ and $\alpha = 0.2$ (upper panel), $\alpha = 0.5$ (middle panel), 
                   $\alpha = 0.8$ (lower panel).
                   The insets show the results when $(k_{{\rm F}} a_{{\rm F}})^{-1}=1$ for the same values of $\alpha$.
                   Here, $\rho$ is in units of $R_{{\rm TF}}$ and densities are in units of 
$(N_{\uparrow} + N_{\downarrow}) / R_{{\rm TF}}^{2}$.}
\end{center}
\end{figure}

(ii) fig.~3 for the corresponding \emph{column} density profiles $n(\rho) = \int \! dz \, n(\mathbf{\rho},z)$ vs.
     $\rho = \sqrt{x^{2} + y^{2}}$.
     Phase separation appears now less visible even for $(k_ {F} a_{{\rm F}})^{-1}=1$, since 
     $n(\rho)$ leaks toward $\mathbf{\rho}=0$ where it acquires a finite value.

\begin{figure}
\begin{center}
\includegraphics*[width=7.5cm]{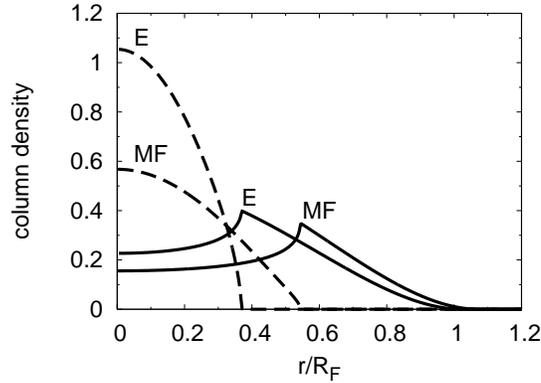}
\caption{Column density profiles $\delta n(\rho)$ (full lines) and $n_{0}(\rho)$ (broken lines) 
                   vs.~$\rho$ when $(k_{{\rm F}} a_{{\rm F}})^{-1}=1$ and $\alpha = 0.5$, as obtained with the exact (E) and 
                   mean-field (MF)
                   values of the scattering lengths. 
                   Units are the same as in fig.~3.}
\end{center}
\end{figure}

(iii) fig.~4 for the comparison of the column density profiles when $(k_{{\rm F}} a_{{\rm F}})^{-1}=1$ and $\alpha = 0.4$, as obtained 
      with the exact and mean-field values of the scattering lengths. 
      The use of the exact values of the scattering lengths results in density profiles which are more 
      compressed toward $\rho = 0$. 
      This feature can be conveniently characterized by the values of the \emph{critical radius} $R_{c}$ where the condensate vanishes.

\begin{figure}
\begin{center}
\includegraphics*[width=13.5cm]{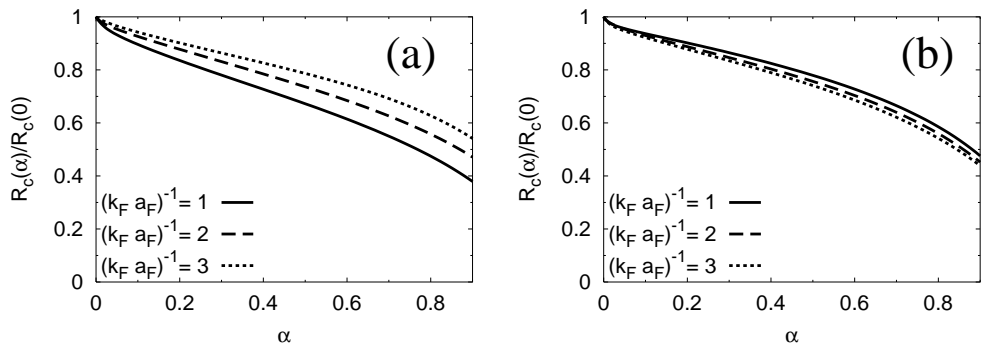}
\caption{Critical radius $R_{c}$ of the condensate density vs.~the asymmetry parameter $\alpha$ (normalized to the
                   value $R_{c}(0)$ at $\alpha=0$) for three coupling values,
                   obtained by using the exact (a) or mean-field (b) values of the scattering lengths.
                   [Adapted from ref.~\cite{PS-2006}.]}
\end{center}
\end{figure}

(iv) fig.~5 for the critical radius $R_{c}$ of the condensate density, which is plotted vs.~$\alpha$ for different
     couplings and using the exact (a) or mean-field (b) values of the scattering lengths.
     This quantity identifies also the position of the maximum of the density of excess fermions.
     Note the inverted sequence of the curves corresponding to the different values of $(k_{{\rm F}} a_{{\rm F}})^{-1}$, as calculated
     with the exact or with the mean-field values of the scattering lengths.
     This feature could be subject to experimental verification.

\begin{figure}
\begin{center}
\includegraphics*[width=7.5cm]{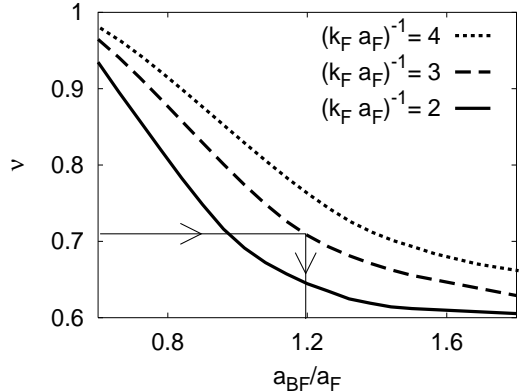}
\caption{Ratio $\nu$ between the column density of excess fermions at the center of the trap and its maximum value 
                   at $R_{c}$ vs. the ratio $a_{{\rm BF}}/a_{{\rm F}}$, for $\alpha=0.5$ and three coupling values.
                   [Adapted from ref.~\cite{PS-2006}.]}
\end{center}
\end{figure}

(v) fig.~6 for the ratio $\nu$ between the column density of the excess fermions at the center of the trap and its maximum 
    value at the critical radius $R_{c}$ vs. $a_{{\rm BF}}/a_{{\rm F}}$ for $\alpha=0.5$ and different couplings.
    The use of column density to obtain the ratio $\nu$ stems from our finding that column density profiles have a more marked 
    dependence on $a_{{\rm BF}}$.
    Indeed, the marked dependence on $a_{{\rm BF}}$ should make it possible to extract the expected value $1.18$ of $a_{{\rm BF}}/a_{{\rm F}}$
    from the experimental data, using the plots of fig.~6 as \emph{calibration curves}.
    This procedure is indicated schematically in the figure.
   
Note that, although the results presented in this Section have been obtained for an isotropic (spherical) trap, they can be also utilized for an anisotropic 
(ellipsoidal) trap, as indicated in Appendix A.

Comparison with the experimental results by the MIT \cite{Ketterle-2006} and Rice \cite{Hulet-2006} groups for the density
profiles ot the two imbalanced fermion species can be established at this point, at least as far as the BEC side of the
crossover region is concerned.
In this regime and at low temperatures, the experiments indicate that the dominant effect of density imbalance is to phase
separate the system in two components, with an inner superfluid region where fermions of different species balance each other
and an outer normal region where the excess fermions reside.
This effect was correctly reported for the first time in ref.~\cite{PS-2006} and it is clearly evidenced in the above figures.

\section{Extension to vortices (rotating frame)}

Unambiguous detection of the superfluid phase for trapped Fermi atoms with balanced populations has been 
achieved by the observation of vortex lattices when the system is put into \emph{rotation} \cite{Ketterle-2005}.
By a similar token, in the presence of density imbalance between the two fermionic species, detection of the superfluid 
region has relied on the observation of vortices in that region \cite{Ketterle-2006}.

To deal with vortex patterns originating from fermions in the presence of density imbalance, we extend the previous
treatment holding on the BEC side of the crossover in the following way.
We begin by noting that the two coupled equations (\ref{local-gap-equation}) and (\ref{density-excess-fermions}) for the 
condensate wave function $\Phi(\mathbf{r})$ and the density of excess fermions $\delta n(\mathbf{r})$ could alternatively
be derived from a suitable \emph{energy functional} $E[\Phi,\delta n]$.
When the system is further put into rotation, this energy functional reads:
\begin{eqnarray}
 E [\Phi,\delta n] &=&  Z \, \int \! dx dy \, \left\{ \frac{1}{4m} \, |\nabla \Phi (\mathbf{r})|^{2}
\, + \, m \omega_{r}^2 \, r^{2} |\Phi (\mathbf{r})|^{2}                                        
+  \frac{\pi a_{{\rm B}}}{m} \, |\Phi (\mathbf{r})|^{4}    
\,  \right. \nonumber \\                             
 &-& \, \Omega \, \Phi (\mathbf{r})^{*} L_{z} \Phi (\mathbf{r})  +   \frac{6^{5/3} \pi^{4/3}}{20 m} \, \delta n(\mathbf{r})^{5/3} 
\, + \, \frac{1}{2} m \omega_{r}^2 \, r^{2} \delta n(\mathbf{r})\nonumber \\                                               
&+&  \left. \frac{3 \pi a_{{\rm BF}}}{m} |\Phi (\mathbf{r})|^{2} \delta n(\mathbf{r})    
\, - \, \frac{1}{2} m \Omega^{2} r^{2} \, \delta n(\mathbf{r}) \right\}\,\, .   \label{energy-functional-vortices}      
\end{eqnarray}
Here, $\Omega$ is the rotation frequency, $L_{z}$ the angular momentum operator of the composite bosons, $\mathbf{r}=(x,y,0)$, 
while for calculation convenience the original ellipsoidal trap has been replaced by a cylinder of height $Z$ with a
harmonic radial potential.
Note that in eq.~(\ref{energy-functional-vortices}) the excess fermions in the rotating frame have been treated within a 
semi-classical description.
Let, in fact, 
\begin{equation}
n_{R}(\mathbf{r}) \, = \, \int \! \frac{d \mathbf{p}}{(2 \pi \hbar)^{3}} \,\,
\frac{1}{e^{ \beta [\epsilon_{R}(\mathbf{p},\mathbf{r}) \, - \, \mu] } \, + \, 1}   \label{generic-semiclassical-fermion-density}
\end{equation}
be the semi-classical fermionic distribution in the rotating ($R$) frame, where
$\epsilon_{R}(\mathbf{p},\mathbf{r}) = \frac{\mathbf{p}^{2}}{2 m} + U(\mathbf{r}) - \mathbf{\Omega} \cdot \mathbf{l}$ the
associated semi-classical dispersion with angular momentum $\mathbf{l}$.
Since the last expression can be manipulated as follows
\begin{eqnarray}
\epsilon_{R}(\mathbf{p},\mathbf{r}) 
& = & \frac{\mathbf{p}^{2}}{2 m} \, + \, U(\mathbf{r}) \, - \, \mathbf{p} \cdot ( \mathbf{\Omega} \times \mathbf{r} )  \nonumber \\
& = & \frac{\mathbf{p'}^{2}}{2 m} \, + \, U(\mathbf{r}) \, - \, \frac{1}{2} m ( \mathbf{\Omega} \times \mathbf{r} )^{2} 
                                                                                                     \label{semiclassical-energy}
\end{eqnarray}
where  $\mathbf{p'} = \mathbf{p} - m \mathbf{\Omega} \times \mathbf{r}$ in the last line, 
eq.~(\ref{generic-semiclassical-fermion-density}) can eventually be cast in the form:
\begin{equation}
n_{R}(\mathbf{r}) \, = \, \int \! \frac{d \mathbf{p'}}{(2 \pi \hbar)^{3}} \,\,
f \left( \frac{\mathbf{p'}^{2}}{2 m} \, + \, U(\mathbf{r}) \, - \, \frac{1}{2} m ( \mathbf{\Omega} \times \mathbf{r} )^{2} \right) 
                                                                                 \label{specific-semiclassical-fermion-density}
\end{equation}
which agrees with the expression for $\delta n(\mathbf{r})$ one obtains from eq.~(\ref{energy-functional-vortices}).

\begin{figure}
\begin{center}
\includegraphics*[width=14.5cm]{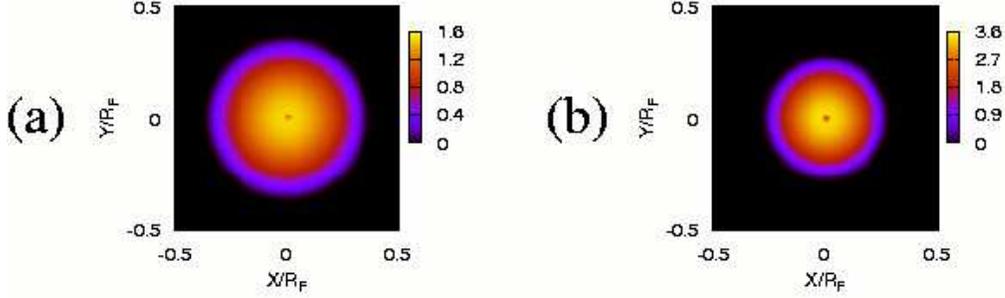}
\caption{Pattern of an isolated vortex in the balanced situation ($\alpha=0$) at the lower critical frequency 
                   for coupling values (a) $(k_{{\rm F}} a_{{\rm F}})^{-1}=1$ and (b) $(k_{{\rm F}} a_{{\rm F}})^{-1}=4$.}
\end{center}
\end{figure}

\begin{figure}
\begin{center}
\includegraphics*[width=14.5cm]{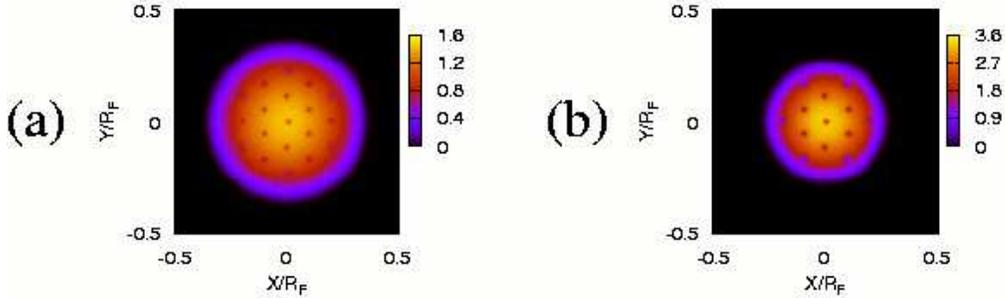}
\caption{Triangular-shape vortex lattice in the balanced situation ($\alpha=0$) when $\Omega = 19 Hz$ and for 
                   coupling values (a) $(k_{{\rm F}} a_{{\rm F}})^{-1}=1$ and (b) $(k_{{\rm F}} a_{{\rm F}})^{-1}=4$.}
\end{center}
\end{figure}

\begin{figure}
\begin{center}
\includegraphics*[width=14.5cm]{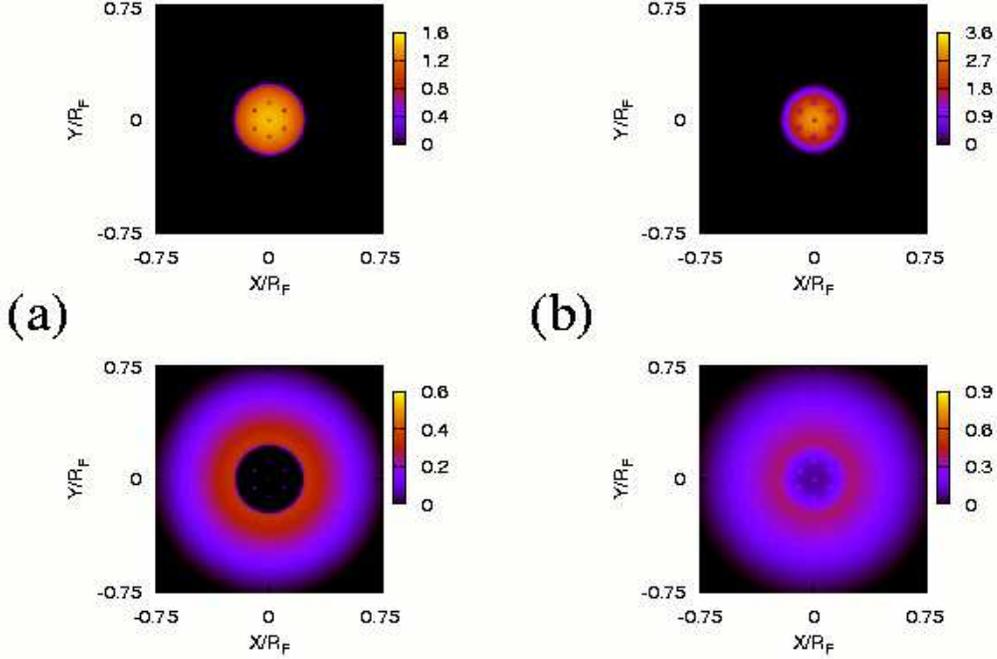}
\caption{Triangular-shape vortex lattice in the imbalanced situation with $\alpha=0.4$, $\Omega = 19 Hz$, 
                   and coupling values (a) $(k_{{\rm F}} a_{{\rm F}})^{-1}=1$ and (b) $(k_{{\rm F}} a_{{\rm F}})^{-1}=4$.
                   The upper (lower) plots refer to the density of condensed composite bosons (excess fermions).}
\end{center}
\end{figure}

Sticking to this equation, one has the choice of \emph{either} solving the associated coupled differential equations for 
$\Phi(\mathbf{r})$ and $\delta n(\mathbf{r})$ \emph{or} minimizing directly $E[\Phi,\delta n]$ by varying $\Phi(\mathbf{r})$ 
and $\delta n(\mathbf{r})$.
In practice, a ``mixed'' approach may sometimes be preferred \cite{SPS-2006}.
Note that in the present treatment the Thomas-Fermi (LDA) approximation is altogether avoided.
[We have also verified that this refinement leads only to minor changes for the density profiles presented in Section 5 in the 
absence of rotation.]

We present numerical results for the parameters of the MIT trap \cite{Ketterle-2006} for which $N = 2 \times 10^{6}$, 
$\omega_{\perp} = 57 Hz$, and $\omega_{z} = 23 Hz$.
We begin by showing in fig.~7 the pattern of an isolated vortex in the balanced ($\alpha=0$) situation at the lower critical 
frequency $\Omega_{c_{1}}$ when the first vortex enters the system, for the two cases
$(k_{{\rm F}} a_{{\rm F}})^{-1}=1$ and $\Omega_{c_{1}} = 3.95Hz$ (a), 
and $(k_{{\rm F}} a_{{\rm F}})^{-1}=4$ and $\Omega_{c_{1}} = 6.11Hz$ (b).
When the rotation frequency increases further, this pattern evolves in a triangular-shape vortex lattice shown in fig.~8 
when $\Omega = 19 Hz$ for the same coupling values as in fig.~7.
Such triangular pattern persists in the imbalanced situation, as shown in fig.~9 for $\alpha=0.4$ and the same frequency 
$\Omega = 19 Hz$, when (a) $(k_{{\rm F}} a_{{\rm F}})^{-1}=1$ and (b) $(k_{{\rm F}} a_{{\rm F}})^{-1}=4$.
In fig.~9 the \emph{upper} plots picture the density of condensed composite bosons while the \emph{lower} plots refer to 
excess fermions.
Note that the excess fermions tend to pile up in the vortex cores, taking thus advantage of the fact that the 
density of composite bosons is depressed there.
Further study along these lines is in progress \cite{SPS-2006}.

\section{Perspectives and open problems}

In this contribution, we have presented a theoretical account of the effects produced by imbalancing the populations 
of trapped fermions. 
Special emphasis has been placed to the strong-coupling (BEC) limit, where theoretical studies can be extended beyond 
BCS mean field by exploiting the diluteness condition of the system. 
Several effects were, however, left out from our analysis. 

First of all, it appears relevant to combine density imbalance with \emph{mass imbalance} of the two fermionic species, 
which may act to promote still novel phenomena in the system and it is being subject to experimental investigations at present.

The present approach was confined to the BEC region where $(k_{{\rm F}}\, a_{{\rm F}})^{-1} \gapprox +1$.
As the experiments \cite{Ketterle-2006,Hulet-2006} actually span the whole crossover region $-1 \lapprox (k_{{\rm F}}\, a_{{\rm F}})^{-1} \lapprox +1$,
theoretical calculations should cover this region, too.
In this context, a mean-field approach could readily produce a qualitative overview of the effects taking place across the crossover
region, including both density and mass imbalance.
As in the case of the BEC region, however, mean-field calculations may lead to quantitative incorrect results.
For instance, at unitarity and for $T \simeq 0$, mean-field calculations \cite{Duan-Yip-2006} result in the value
$\alpha_{c}^{MF} \simeq 1$ for the critical asymmetry parameter $\alpha_{c}$ past which the superfluid region disappears
from the system, while the experiment \cite{Ketterle-2006} yields $\alpha_{c}^{exp} \simeq 0.70$.
This is a clear indication that inclusion of \emph{pairing fluctuations} beyond mean field is relevant in the crossover region
to account for the experimental data. 

At higher temperatures, density imbalance may give access to \emph{``precursor'' pairing} (pseudo-gap) effects for the excess 
fermions, which could thus be evidenced in the normal phase \emph{even below} the critical temperature $T_{c}$.
This may result in a strong suppression of thermal fluctuations, leading to a possible detection of an underlying Quantum 
Critical Point.

Finally, we mention that a compelling test for theories would result from calculating the dependence of $T_{c}$ on $\alpha$,
as the experiments were able to determine the temperature of the system directly from the ``tail'' of the density profiles 
of the excess fermions \cite{Ketterle-II-2006}.

\appendix

\section{Appendix A: Mapping of the anisotropic onto the isotropic problem}

The calculations presented in Section 5 refer to an isotropic harmonic potential (spherical trap).
In the experiments with cold Fermi atoms, however, anisotropic harmonic traps are most often used.
The question thus naturally arises to what an extent the theoretical calculations done for a spherical trap
can account for the experimental situation.

The answer to this question relies on the validity of the Thomas-Fermi (LDA) approximation, which applies 
when the number $N$ of trapped atoms is large enough and was explicitly used in the calculations of Section 5. 
[As a matter of fact, when $N$ is large enough, avoiding this approximation leads only to minor changes of the 
density profiles, as mentioned in Section 6.]

This is because, within LDA, physical quantities acquire an $\mathbf{r}-$dependence \emph{only through}
the local chemical potential
\begin{equation}
\mu(x,y,z) \, = \, \mu \, - \, \frac{1}{2} \, m \, \omega_{x}^{2} \, 
(x^{2} \, + \, \lambda_{y}^{2} y^{2} \, + \, \lambda_{z}^{2} z^{2})   \,\,\, .                 \label{local-chemical-potential} 
\end{equation}
Here, the anisotropy of the harmonic potential is specified by the parameters $\lambda_{y}$ and $\lambda_{z}$.
In particular, the density profiles of interest can be expressed as $n(x,y,z) = n[\mu(x,y,z)]$.

To establish the desired mapping with the isotropic case, we introduce an \emph{equivalent isotropic problem} with
a total number of atoms $N_{{\rm iso}} = \lambda_{y} \lambda_{z} N$, the harmonic frequency $\omega_{{\rm iso}} = \omega_{x}$,
and the same Fermi energy $E_{{\rm F}}^{{\rm iso}} = E_{{\rm F}} = \left( 3 \omega_{x} \omega_{y} \omega_{z} N \right)^{1/3}$ of the
original anisotropic problem.
To determine the corresponding chemical potential $\mu_{{\rm iso}}$ for this problem, we argue that the ratio $\mu_{{\rm iso}}/E_{{\rm F}}$ 
must be a universal function of the dimensionless quantities $(k_{{\rm F}} a_{{\rm F}})^{-1}$ and $T/E_{{\rm F}}$ (where, by definition,
$k_{{\rm F}} = \sqrt{2 m E_{{\rm F}}}$).
This implies that the ratio $\mu_{{\rm iso}}/E_{{\rm F}}$ can depend on $N$ only via $E_{{\rm F}}$.
As a consequence, $\mu_{{\rm iso}}$ of the equivalent isotropic problem must coincide with the chemical potential $\mu$ of the 
original anisotropic problem.

For the density profile $n(x,y,z)$ of the original anisotropic problem, this equivalence eventually implies that it can be
expressed in terms of the density profile $n_{{\rm iso}}(x,Y,Z)$ of the equivalent isotropic problem as follows:
\begin{equation}
n(x,y,z) \, = \, n_{{\rm iso}} \left[ \mu_{{\rm iso}} \, - \, \frac{1}{2} m \omega_{{\rm iso}}^{2} (x^{2} + Y^{2} + Z^{2}) \right]     \label{n-n-iso}
\end{equation}
where $Y = \lambda_{y} y$ and $Z = \lambda_{z} z$.

\section{Appendix B: Axial density profiles}

The density profiles reported in Section 5 were obtained within the Thomas-Fermi (LDA) approximation.
This approximation tends to sharpen the density profiles at their edges, but is otherwise appropriate when the number of 
trapped atoms is large enough.
From those results we have also concluded that the dominant effect of density imbalance is to phase separate the system into 
two components, an inner superfluid region and an outer normal region with excess fermions only.

In this context, an argument was proposed by De Silva and Mueller \cite{DSM-2006} to evidence the occurrence of phase 
separation in the density profiles when the trapping potential is \emph{harmonic}.
Their argument relies on the validity of the LDA approximation and applies to the \emph{axial} density $n_{{\rm A}}(z)$, 
obtained from the density through a double integration: 
\begin{equation}
n_{{\rm A}}(z) =  \int \! dx \, dy \, n(x,y,z)                                            
         =  \pi \, \int_{0}^{\infty} \! \, d \rho^{2} \, n(\rho^{2} \, + \, z^{2})  
         =  \pi \, \int_{z^{2}}^{\infty} \! \, d \zeta \, n(\zeta)  
\label{axial-density-profile} 
\end{equation}
where $\zeta = \rho^{2} + z^{2}$ and $n(\zeta) \geq 0$. 

\begin{figure}
\begin{center}
\includegraphics*[width=6.5cm]{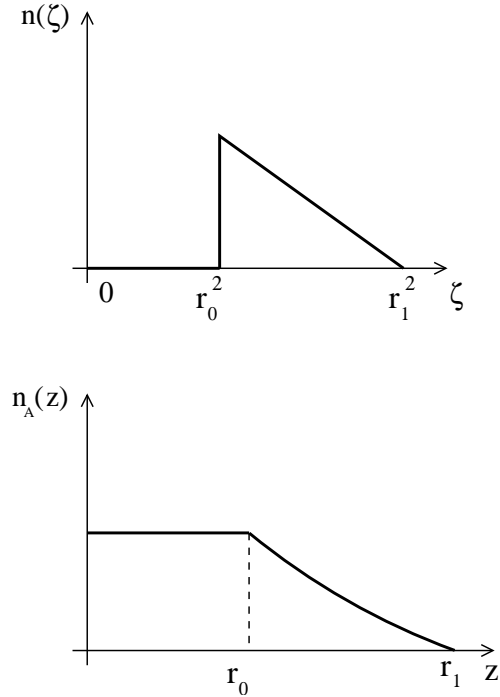}
\caption{Schematic plot of the density $n(\zeta)$ with a hole in the core region and of the corresponding 
                    axial density $n_{{\rm A}}(z)$ with a flat behavior in the core.}
\end{center}
\end{figure}
From the last line of eq.~(\ref{axial-density-profile}) one readily concludes that $n_{{\rm A}}(z)$ is a \emph{decreasing} 
function of $z$.
In addition, when $n(x,y,z)$ presents a ``hole'' in the core region (say, for $r < r_{0}$), $n_{{\rm A}}(z)$ is \emph{constant} 
in that region.
This situation is pictured in fig.~10, where schematic plots of the radial density $n(\zeta)$ with a hole in the core region 
(upper panel) and of the corresponding axial density $n_{{\rm A}}(z)$ with a \emph{flat} behavior in the core (lower panel) 
are shown. 
This situation is a fingerprint for the occurrence of phase separation and applies to the density profiles shown 
in the insets of fig.~2.

\acknowledgments

We thank Stefano Simonucci, who performed the numerical calculations on vortices and prepared figs.~7-9. This work was  
partially supported by the Italian MIUR (contract Cofin-2005 ``Ultracold Fermi Gases and Optical Lattices'').



\end{document}